\documentclass[twocolumn,showpacs,preprint,preprintnumbers,superscriptaddress,aps,amsmath,amssymb,10pt]{revtex4-1}% Physical Review B

\usepackage{graphicx}
\usepackage{dcolumn}
\usepackage{bm}
\usepackage{units}

\begin{document}

\title{Plastic yielding in nanocrystalline Pd-Au alloys mimics universal behavior of metallic glasses}

\author{A.~Leibner}
\email[]{leibner@nano.uni-saarland.de}
\affiliation{Experimentalphysik, Universit\"at des Saarlandes, Saarbr\"ucken, Germany}
\author{C.~Braun}
\affiliation{Experimentalphysik, Universit\"at des Saarlandes, Saarbr\"ucken, Germany}
\author{J.~Heppe}
\affiliation{Experimentalphysik, Universit\"at des Saarlandes, Saarbr\"ucken, Germany}
\author{M.~Grewer}
\affiliation{Experimentalphysik, Universit\"at des Saarlandes, Saarbr\"ucken, Germany}
\author{R.~Birringer}
\affiliation{Experimentalphysik, Universit\"at des Saarlandes, Saarbr\"ucken, Germany}

\date{\today}

\begin{abstract}
We studied solid solution effects on the mechanical properties of nanocrystalline (NC) $\mathrm{Pd}_{\mathrm{100-x}} \mathrm{Au}_{\mathrm{x}}$ alloys ($0 \leq \mathrm{x} < \unit[50]{at.\%}$) at the low end of the nanoscale. Concentration has been used as control parameter to tune material properties (elastic moduli, Burgers vector, stacking fault energies) at basically unaltered microstructure (grain size $D\approx \unit[10]{nm}$). In stark contrast to coarse grained fcc alloys, we observe solid solution softening for increasing Au-content. The available predictions from models and theories taking explicitly into account the effect of the nanoscale microstructure on the concentration-dependent shear strength have been disproved without exception. As a consequence, it is implied that dislocation activity contributes only marginally to strength. In fact, we find a linear correlation between shear strength and shear modulus which quantitatively agrees with the universal behavior of metallic glasses discovered by Johnson and Samwer [W.L. Johnson and K. Samwer, PRL 95, 195501 (2005)]. %\\[2ex]

\end{abstract}


\maketitle

% INTRODUCTION %%%%%%%%%%%%%%%%%%%%%
\section{introduction} %%%%%%%%%%%%%%%%%%%%%
Over the past two decades, it has been well established that decreasing the grain size $D$ of polycrystalline metals into the nanometer regime, $D < \unit[100]{nm}$, results in e.g. a substantial increase of strength \cite{Meyers2006}, improved fatigue \cite{Boyce2010} as well as wear resistance \cite{Rupert2013}. Gaining insight into the physics of the underlying deformation mechanisms has motivated intense research efforts with a focus on studying microstructure-dependent deformation behavior with grain size as prominent control parameter \cite{Dao2007}. In particular, when decreasing the grain size to the lower end of the nanoscale $D \lesssim \unit[10]{nm} $, it has been argued that \textit{intragranular} crystal plasticity becomes to a large proportion replaced by \textit{intergranular} plasticity, i.e. deformation processes which essentially emerge in the core region of grain boundaries (GBs). Indeed, computer simulations and experiments unraveled a variety of modes of plastic deformation that are mediated by GBs: GB slip and sliding \cite{Vo2008, VanSwygenhoven2001, Weissmuller2011}, grain rotation \cite{Legros2008, Ma2004, Shan2004} that may even lead to grain coalescence but is also an integral part of stress-driven GB migration (SDGBM) \cite{Cahn2004, Cahn2006, Legros2008} as well as of shear transformations (STs) \cite{Lund2005, Argon2006}. The latter involve shuffling or flipping of groups of atoms and may act as flow defect operating in the core region of GBs, thus playing a similar role in a disordered proximity as dislocations do in a crystalline environment. Moreover, GB facets or ledges and triple junctions act as stress concentrators thereby effectively reducing the barrier for partial dislocation nucleation and emission \cite{Gu2011, Asaro2005}. Because of the complex interplay of disparate mechanisms, operating either sequentially or simultaneously, it is still a controversial issue which role they play in responding to the intrinsic stress field and which share to overall strain propagation is carried by them. 

To improve our understanding of how different plasticity mechanisms interact and contribute to strain propagation on the nanoscale, we study solid solution alloying and its effect on the strength of the material. Fortunately, the Pd-Au alloy system, which is fully miscible and has a negligibly small tendency to segregation, enables to prepare any Au-concentration at basically fixed grain size of $\approx \unit[10]{nm}$. It so becomes feasible to \textit{gradually tune material parameters } (lattice parameter, Burgers vector, elastic moduli, stacking fault energy, GB energy) and explore their influence on plastic deformation behavior without changing the character of microstructure (grain size, texture). 

Regarding deformation mechanisms, recent studies on NC Pd$_{90}$Au$_{10}$ have unraveled that plastic deformation is governed by shear shuffling (STs) at/along GBs while dislocation activity more likely plays a minor role \cite{Grewer2014, Grewer2014arxiv}. Nevertheless, it is an open problem whether or not an increase of Au-concentration involving a concomitant change of material parameters will lead to a change of the dominant deformation mechanism(s) operating at the nanoscale. In particular, the Pd-Au alloy system exhibits a high stacking fault energy $\approx \unit[180]{mJm^{-2}}$ in the Pd-rich alloys and a low stacking fault energy $\approx \unit[50]{mJm^{-2}}$ on the Au-rich side \cite{Schaefer2011, Jin2011} and, therefore, we expect an increasing propensity for partial dislocation emission from GBs that goes in parallel with lowering stacking fault energies. To explore this scenario, we compare the evolution of strength of coarse-grained (CG) and NC 
$ \mathrm{Pd}_{\mathrm{100-x}} \mathrm{Au} _{\mathrm{x}}$ alloys $(0 \leq \mathrm{x} < 50$) with the predictions of available theories of solid solution strengthening, relying without exception on dislocation physics. 

% Solid Solution Strengthening: Theory and Models %%%%%%%%%%%%%%%%%%%%%
\section{Solid Solution Strengthening: Theory and Models} %%%%%%%%%%%%%%%%%%%%
Traditional solid solution strengthening theories rely on the idea that solute atoms, which modify the elastic energy of a dislocation, act as obstacles to dislocation motion in a crystalline environment. To characterize the dependence of flow stress or hardness on composition $c$, different models have been suggested which predict linear or power-law ($ c^{1/2}, c^{2/3}$) increase of flow stress with concentration \cite{Haasen1973, Fleischer1964, Labusch1972}. As will be shown later, solid solution strengthening of CG Pd-Au alloys, which serve as reference system, can be sufficiently good described by the classical Fleischer theory \cite{Fleischer1964}. Here, the increase of shear strength is given by 
\begin{align}
	& \Delta \tau = A \cdot G \cdot \varepsilon^{3/2} \cdot c^{1/2} \label{eq:Fleischer} , \\[1.5ex] 
	& \varepsilon = \left| \frac{\varepsilon_G}{1+1/2 \cdot \left| \varepsilon_G \right|} - 3 \varepsilon_b \right| \label{eq:Fleisch} ,
\end{align}%
where $A$ is a material constant, $G$ the shear modulus of the solvent, $b$ the burgers vector of the solvent, and $c$ the concentration of solute atoms. The increase in shear strength $\Delta \tau$ is primarily a consequence of the \textit{local} mismatch in shear modulus ($\varepsilon_G = \frac{1}{G} \frac{\partial G}{\partial c}$ ) and size ($\varepsilon_b = \frac{1}{b} \frac{\partial b}{\partial c}$). Overall, this mismatch results in an effective barrier for dislocation glide. 

Clearly, at the low end of the nanoscale the abundance of GBs, the volume fraction of which scales as $1/D$, has to be taken into account. Rupert et al. \cite{Rupert2011} adapted Fleischer´s model to NC metals by adding two terms which allow for strength enhancement as well as softening. The first term comprises dislocation pinning at GBs and the second term renormalization of this pinning potency by considering the \textit{global} changes of elastic properties and Burgers vector of the abutting crystallites which are sensed when dislocations are bowing across them. The total shear strength of NC alloys has been expressed as 
\begin{equation}
	\begin{aligned}
\tau_{nc} = & \frac{Gb}{D} + A \cdot G \cdot \varepsilon^{3/2} \cdot c^{1/2}\\[1.5ex] 
 + & \frac{Gb}{D}\left(\frac{1}{G}\frac{\partial G}{\partial c} + \frac{1}{b} \frac{\partial b}{\partial c} \right) \cdot c ,
\label{eq:Rupert}
	\end{aligned}
\end{equation}%
where $D$ is the grain size of the pure metal and all other symbols have the meaning already introduced above. The critical shear strength has three contributions: grain size induced hardening, classical Fleischer hardening, and hardening or softening related to the global effects of solute addition ($\partial G / \partial c, \partial b / \partial c$) which are linear in $c$. In fact, the derived relation is capable of predicting solid solution softening whenever $G$ decreases sufficiently strong with increasing $c$. The experimentally observed softening in NC Ni-Cu alloys \cite{Shen1995} and NC Fe-Cu alloys \cite{Fecht1995} could be well-described by Eq. \ref{eq:Rupert}. But also solution strengthening observed in NC Ni-W alloys \cite{Rupert2011} that revealed a rather linear increase in hardness could be equally well-described by Eq. \ref{eq:Rupert}. 

To point to the significance of the stacking fault energy in controlling strength in NC metals, Asaro et al. \cite{Asaro2005} proposed that emission of partial dislocations from GBs, which basically traverse the entire grain at $D \approx \unit[10]{nm}$, may determine the strength of the NC material. In this scenario, preexistence of dislocations (partial or perfect segments) in GBs is assumed. The shear stress resolved along the direction of the lead partial dislocation $\tau_{\mathrm{ped}}$ is given as 
\begin {equation}
\frac{\tau_{\mathrm{ped}}}{G} = \frac{\gamma_{\mathrm{isf}}}{G b} + \frac{1}{3} \frac{b}{D} ,
\label{eq:Asaropartial}
\end{equation}%
where $\gamma_{\mathrm{isf}}$ is the intrinsic (or stable) stacking fault energy which controls the equilibrium spacing of partial dislocations in an unstressed crystal. Alternatively, Asaro et al. also considered emission of partial dislocations from locations of stress concentrations at GBs such as GB facets or triple junctions. The required shear stress to activate such a source is given by 
\begin{equation}
\frac{\tau_{\mathrm{sc}}}{G} = \sqrt{\frac{8}{\pi} \frac{\gamma_{\mathrm{usf}}}{G (1-\nu)}\frac{1}{D}} ,
\label{eq:stressconc}
\end{equation}%
where $\gamma_{\mathrm{usf}}$ is the unstable stacking fault energy and it has been assumed that $D$ is twice the size of a GB facet. Both models have been devoted to pure metals. Nevertheless, we may presume that the effect of alloying is basically captured by the concentration-dependent stacking fault energies, shear moduli and Burgers vectors. The bowing of partial dislocations across a medium which is modified by substitutional solute atoms is however not taken into account here. For the sake of argument, we assume that the concentration-dependent and markedly varying stacking fault energy dominates the deformation behavior. 

As a result of the above survey, validation of solid solution strengthening theories by experiment requires to evaluate how shear modulus $G$, Poisson ratio $\nu$, Burgers vector $b$, which for fcc metals has a direct relation to lattice parameter $a$, grain size $D$ and stable/intrinsic $\gamma_{\mathrm{isf}}$ as well as unstable $\gamma_{\mathrm{usf}}$ stacking fault energy vary with Au-concentration. Knowledge of the full set of material parameters then allows one to compare theory and experiment on a quantitative basis. Except for stacking fault energies, we have determined all relevant material parameters by experiment. Values for the stable and unstable stacking fault energies of the Pd-Au system are provided by MD simulations of Sch\"afer et al. \cite{Schaefer2011} as function of concentration as well as ab initio calculations of Jin et al. \cite{Jin2011} for the pure metals Pd and Au. Linear variation of stacking fault energies with composition across the whole composition range of Pd-Ag alloys were found in experiment \cite{Harris1966} and ab initio electronic structure calculations \cite{Crampin1993}. To allow for comparison, we also linearly extrapolate the values from \cite{Jin2011}. Relevant stacking fault energies are summarized in Fig. \ref{fig3}. In what follows, we discuss sample preparation and how we extract material parameters by utilizing X-ray diffraction, ultrasound, and Vickers-microhardness indentation tests. 

% Preparation and Methodology %%%%%%%%%%%%%%%%%%%%%
\section{Preparation and Methodology} %%%%%%%%%%%%%%%%%%%%
The binary NC Pd-Au samples, with Au-concentration between 0 and 50 atomic\%, were prepared by inert gas condensation (IGC) and compaction at 1.8 GPa \cite{Birringer1989} to obtain disc-shaped samples with a diameter of \unit[8]{mm} and a thickness of about \unit[1]{mm}. IGC-prepared samples have a random texture and lognormally distributed equiaxed grains \cite{Skrotzki2013, Krill1998} with a volume-weighted average grain size $D_{\mathrm{vol}} \approx \unit[10]{nm}$. The latter was determined using Klug and Alexander's \cite{Klug1974,Markmann2008} modified Williamson-Hall technique applied to Bragg-peak broadening of X-ray diffraction diagrams. Lattice parameters have been determined from a Nelson-Riley \cite{Kasper1972} plot of \{hkl\}-dependent Bragg peak positions. All diffraction experiments were performed on a laboratory diffractometer (PANalytical XPert Pro) operated in Bragg-Brentano focusing geometry and $\theta - \theta$ mode. The composition of as prepared specimens was determined by EDX (EDAX TSL Trident system) in a SEM (JEOL F 7000). CG Pd-Au samples were prepared by arc melting, cold deformation to a disc, and subsequent annealing at $\unit[900]{^{\circ} C}$ which causes primary and secondary recrystallization to end in an average grain size of $\approx \unit[100]{\mu m}$. Alternatively, we annealed NC Pd-Au specimens at $\unit[400]{^{\circ}C}$ to induce curvature-driven grain-growth resulting in an average grain size of $\approx \unit[100]{\mu m}$. CG specimens were characterized in analogy to the NC samples except grain size was determined by electron backscatter diffraction (EBSD) in the SEM. 

All specimens were coupled to a \unit[20]{MHz} ultrasonic transducer (Panametrics V2173), capable to simultaneously transmit longitudinal and transverse waves. The ultrasonic transducer was connected to a LeCroy WaveRunner 6051 digital oscilloscope which allowed us, by applying the pulse-echo overlap method \cite{Papadakis1990}, to extract time-of-flight times of the waves. The velocities of longitudinal and transverse (shear) waves are given by the ratios of two times the specimen thickness over the respective time-of-flight. For a quasi-isotropic material and assuming linear elasticity, the following relations hold between the scalar shear (G) and Young's (E) moduli and the longitudinal and transverse sound velocities, $v_{\mathrm{l}}$ and $v_{\mathrm{s}}$ \cite{Marder1960} 
\begin{align}
G =& \rho\cdot v_{\mathrm{s}}^{2} \label{Shear} , \\[1.5ex]
E =& \rho\cdot v_{\mathrm{s}}^{2} \cdot \frac{3v_{\mathrm{l}}^{2}-4v_{\mathrm{s}}^{2}}{v_{\mathrm{l}}^{2}-v_{\mathrm{s}}^{2}} , \label{Young}
\end{align}%
where $\rho$ is the sample density. The overall density of NC-materials is reduced compared to the density of their CG counterparts. This is due to the fact that the core regions of GBs carry excess volume \cite{Birringer1995, Shen2008} resulting from atomic site mismatch that is created when two differently oriented crystal lattices meet along a common interface. A few percent porosity due to processing is a second source of density reduction. The overall density of NC materials can be determined with high accuracy using the method of Archimedes in conjunction with a microbalance (reference media: air and diethyl phthalate) \cite{Hoffmann1996}. In Appendix A, we discuss how excess free volume and porosity can be discriminated likewise how measured densities can be corrected for porosity. 

It remains to be addressed that Poisson's ratio $\nu$ depends on $E$ and $G$ and assumes the following form
\begin{equation}
\nu = (E-2G)/2G ;
\label{eq:poisson}
\end{equation}%
more details on this matter can be found in \cite{Grewer2011}. Vickers hardness measurements were performed on a Frank Durotest 38151 testing device applying a testing force of $\unit[980]{mN}$ (HV0.1) and averaging over 20 indents per NC sample. Following the pertinent literature \cite{Tabor1951, Barrett1973}, we employed the relation $HV \approx 3 \sigma_y \approx 3(2\tau_y)$ and use microhardness (indent-diagonal $>\unit[20]{\mu m}$) as a measure of shear stress at yielding. 
\begin{figure}[htb!]%
\includegraphics[width=\columnwidth]{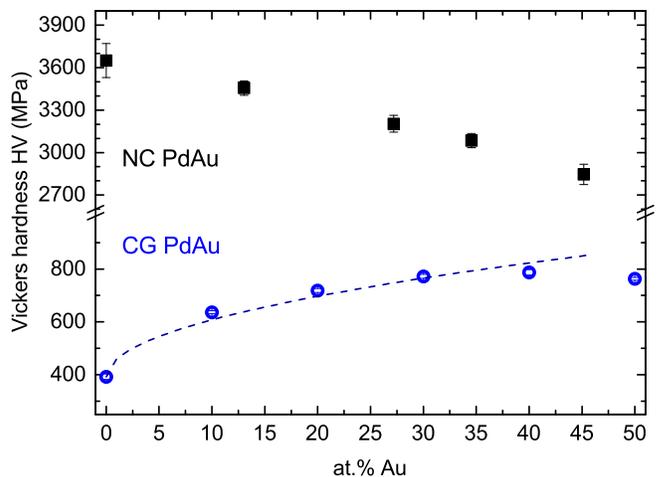}
\caption{(Color online) Vickers hardness HV of nanocrystalline (\textit{black squares}) and coarse grained (\textit{blue circles}) Pd-Au as a function of Au-concentration. Dashed line represents a least-squares fit to the data points based on the prediction from Fleischer's model according Eq. \ref{eq:Fleischer}.}%
\label{fig1}
\end{figure} 

% RESULTS and DISCUSSION %%%%%%%%%%%%%%%%%%%%%
\section{Results and Discussion} %%%%%%%%%%%%%%%%%%%%
In Fig. \ref{fig1} we display the Vickers hardness as a function of Au-concentration for both CG and NC Pd-Au alloys. As expected, \textit{CG Pd-Au ($D \approx \unit[100]{\mu m}$) shows classical solid solution hardening behavior, whereas, the hardness of the NC alloys decreases with rising Au-content.} Before focusing on this fundamental discrepancy, we set the benchmark for comparison by examining whether the CG alloys agree with the predictions from Fleischer's model (Eq. \ref{eq:Fleischer}). 

\begin{figure}[ht]
\includegraphics[width=\columnwidth]{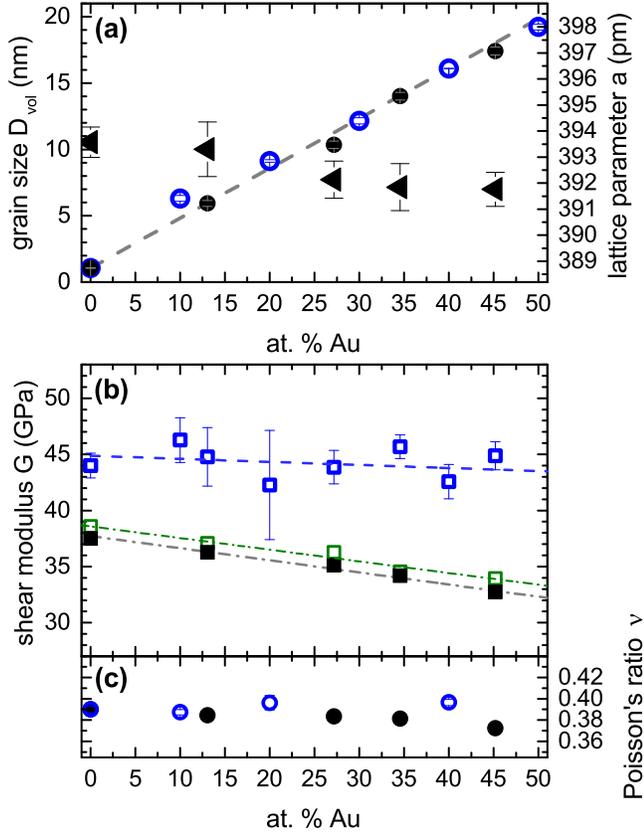}
\caption{(Color online) The following parameters are plotted as function of global Au-concentration: (a) Mean grain size $D_{\mathrm{vol}}$ (\textit{black triangles}) of NC Pd-Au and lattice parameter $a$ of both NC Pd-Au (\textit{black dots}) and coarse-grained samples (\textit{blue circles}). Dashed line connects $a_{\mathrm{Pd}}$ and $a_{\mathrm{Au}}=\unit[407.9]{pm}$ \cite{Smithells} according to Vegard's rule. (b) Shear modulus $G$ of coarse grained (\textit{open blue squares}) and NC Pd-Au.
\textit{Black squares} represent $G$-values not corrected for porosity and the \textit{open green squares} show the related $G$-values but corrected for porosity (for details see Appendix A). Dashed or dashed-dotted lines are linear fits to the data points. (c) Poisson's ratio $\nu$ of CG (\textit{open blue circles}) and NC samples (\textit{black dots}); $\nu$ is basically not affected by porosity.}
\label{fig23}
\end{figure}

The needed material parameters (lattice parameter, shear modulus) are displayed in Fig. \ref{fig23} together with the data for the NC alloys. Clearly, the lattice parameters follow Vegard's rule and it is straightforward to determine $\partial b/\partial c = \unit[(13.32 \pm 0.06) \cdot 10^{-2}]{pm/at.\%}$, where it has been assumed that for full dislocations in a fcc lattice $b = a/\sqrt{2}$, and $b_p = a/\sqrt{6}$ for partial dislocations. Surprisingly, the shear modulus of the CG alloys exhibits only a weak concentration dependence $\partial G/\partial c = \unit[(-0.3 \pm 0.2) \cdot 10^{-3}]{GPa/at.\%}$. With the solvent values for $b = \unit[275.0 \pm 0.1]{pm}$ and $G = \unit[44.0 \pm 0.5]{GPa}$ taken from the pure Pd specimen and using least-squares fitting, we can verify that the Fleischer model (dashed line in Fig.\ref{fig1}) is in good agreement with our experimental data. The parameter $A$ in Eq.1 is a material specific constant which was treated as free fit parameter and has been determined to $A = 0.029 \pm 0.001$. 

To analyze the observed solution softening behavior of NC Pd-Au alloys (Fig. \ref{fig1}) the model of Rupert et al. \cite{Rupert2011} seems predestined to be applied here since the negative slope of $\partial G/ \partial c$ (Fig. \ref{fig23} (b)) is a necessary prerequisite for showing softening. With the grain size $D$ of pure NC Pd taken from Fig. \ref{fig23} (a) and $A = 0.029$, it is straightforward to compute $\tau_{\mathrm{nc}}$ according to Eq. \ref{eq:Rupert}. As displayed in Fig. \ref{fig4}, the model of Rupert et al. neither reveals softening nor is it capable of agreeing with the magnitude of the combined size and alloying effects in the NC Pd-Au alloys. 
 
Concerning this discrepancy, one could argue that the effect of dislocation motion through NC pinning points is better approximated by the grain sizes related to the alloy specimens instead of the slightly larger value of pure NC Pd. But this would even enhance the discrepancy connected with the magnitude of strength without giving rise to softening. Segregation or desegregation of solutes to/from GBs may also be invoked as a source of disagreement. However, as shown in Fig. \ref{fig23} (a), the lattice parameters of CG as well as NC Pd-Au alloys follow the same Vegard rule implying that pronounced Au-segregation to GBs can be ruled out to cause softening. We are not intending to discard the model of Rupert et al., in fact, we suppose that the assumptions made in this model may properly describe the deformation behavior of NC alloys at grain sizes larger than \unit[20 - 30]{nm}. 

\begin{figure}[htb!]%
\includegraphics[width=\columnwidth]{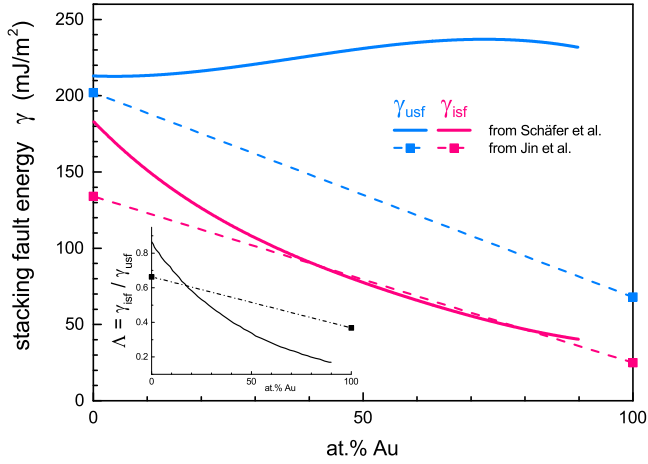}%
\caption{(Color online) Stable (isf) and unstable (usf) stacking fault energies $\gamma$ of $\mathrm{Pd}_{\mathrm{100-x}} \mathrm{Au}_{\mathrm{x}}$. Data (full lines) were taken from Sch\"{a}fer et al. \cite{Schaefer2011}, and values for pure Pd and Au (squares) from Jin et al. \cite{Jin2011}. Broken lines represent linear interpolations between the pure material values from Jin et al. \cite{Jin2011}. Inset: ratio between $\gamma_{\mathrm{isf}}$ and $\gamma_{\mathrm{usf}}$ plotted against Au-concentration. }%
\label{fig3}%
\end{figure}

\begin{figure}%
\includegraphics[width=\columnwidth]{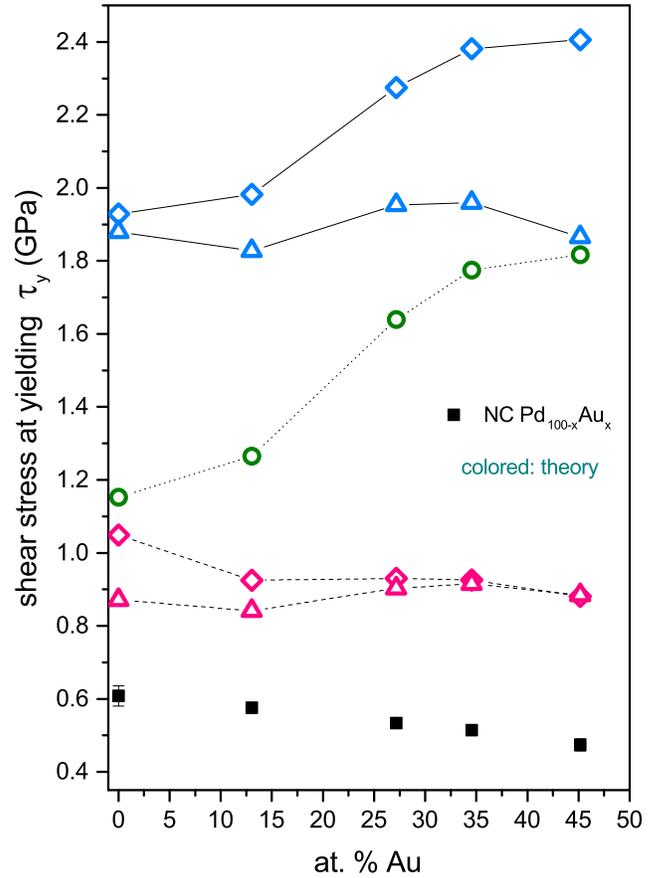}%
\caption{(Color online) Comparison of different theories for limiting shear strength ($\tau_y$) mechanisms and experiment. \textit{Black squares}: shear strength of NC Pd-Au deduced from Vickers hardness measurements using $HV \approx 6\, \tau_y$. \textit{Green circles}: NC pinning model for solid solution strengthening/softening. \textit{Blue symbols/solid lines}: emission of partials from stress concentrations ($\tau_{\mathrm{sc}}$), \textit{red symbols/dashed lines}: emission of partials from preexisting dislocations ($\tau_{\mathrm{ped}}$). \textit{Diamonds} represent data using stacking fault energies from Sch\"{a}fer et al. \cite{Schaefer2011} and \textit{triangles} refer to the straight line extrapolations connecting the stacking fault energy data from Jin et al. \cite{Jin2011}.}
\label{fig4}%
\end{figure}

Eventually, we scrutinize the influence of stacking fault energy on the deformation behavior of NC Pd-Au alloys by referring to Asaro's models. The material parameters shown in Figs. \ref{fig23} and \ref{fig3} have been used as input parameters to Eqs. \ref{eq:Asaropartial} and \ref{eq:stressconc}. When increasing Au-concentration is followed by a decrease in stacking fault energy, one argues that the respective Pd-Au alloy systems become increasingly susceptible to partial dislocation emission, planar slip, fault- and twin formation. To discriminate between twinning and dislocation-mediated slip, Jin et al. \cite{Jin2011} defined a characteristic material measure $\Lambda = \gamma_{\mathrm{isf}}/\gamma_{\mathrm{usf}}$ (see inset Fig. \ref{fig3}) which is correlated with the tendencies to emit partial dislocations, perfect dislocations and twins. Based on a universal scaling law for planar fault energy barriers, they argued that a relatively large value of $\Lambda \approx 0.7$ e.g. of NC $\mathrm{Pd}_{90} \mathrm{Au}_{10}$ (inset Fig. \ref{fig3}) indicates that emission of trailing partials leading to perfect dislocations is favored over twin nucleation. When $\Lambda > 0.8$, it is suggested that twinning can be basically discarded as competitive deformation mode \cite{Jin2011}. It is therefore tempting to assume that partial dislocation emission dominates strain propagation. However, this reasoning is in conflict with recent detailed studies of dislocation activity in NC $\mathrm{Pd}_{90} \mathrm{Au}_{10}$ \cite{Skrotzki2013, Grewer2014arxiv}. Even at high applied strains $\gg 1$, partial- and full dislocation glide has been shown to only marginally contribute to overall strain. This evidence fully agrees with our observations displayed in Fig. \ref{fig4}, reflecting that neither the emission of partials from preexisting dislocation segments (Eq. \ref{eq:Asaropartial}) nor from stress concentrators located in/at GBs (Eq. \ref{eq:stressconc}) are reasonably compatible with the experimental data. Therefore, we conclude that either these models are imperfect or the invoked presuppositions are not met by the NC Pd-Au alloys. 

In fact, the latter argument seems to be valid because detailed studies of thermal activation parameters in NC $\mathrm{Pd}_{90} \mathrm{Au}_{10}$ \cite{Grewer2014} in conjunction with the above mentioned investigations of dislocation activity \cite{Skrotzki2013, Grewer2014arxiv} have unraveled that dislocation scarcity makes room for GB-mediated deformation in NC metals \cite{Grewer2014arxiv}. In particular, shear shuffling or STs have been identified as the major carrier of strain. In other words, in the limit of small grain sizes ($D \lesssim \unit[10]{nm}$), it appears that NC metals mimic glassy behavior. In order to independently verify this idea, we are going to compare the mechanical behavior of NC Pd-Au alloys with the deformation behavior of bulk metallic glasses (BMG). 

Johnson and Samwer \cite{Johnson2005} noted that the shear yield stresses $\tau_{y}$ of metallic glasses at room temperature exhibit universal behavior. Based on compressive yield stress $\sigma_y$ data and using $\tau_y = \sigma_y/2$, they extracted the linear correlation $\tau_y / G = 0.0267 \pm 0.002$ from mechanical testing of more than 30 different metallic glasses. Mechanistically, the elastic-to-plastic transition involves a percolation of STs in space and further deformation increments generate new STs. 

A second aspect of universal behavior relates to the temperature dependence of shear yield stress. Based on a cooperative shear model, they predicted and experimentally verified that temperature dependencies of the shear yield stresses of a large number of metallic glasses fit the universal scaling law 
$\tau_y = \hat{\tau} - \tau(\dot{\gamma}, T) \cdot (T/T_g)^{2/3}$ for temperatures $T \lesssim T_g$, where $T_g$ is the glass transition temperature. The athermal threshold stress $\hat{\tau}$ represents the maximum level of shear resistance as $T \rightarrow \unit[0]{K}$. The term $\tau(\dot{\gamma}, T)$, where $\dot{\gamma}$ is the prescribed strain rate, has been estimated to be very weakly temperature dependent and for typical strain rates of $\unit[10^{-2} - 10^{-4}]{s^{-1}}$ can be approximated as constant $\tau(\dot{\gamma}, T) = (0.016 \pm 0.002) \, G$. 

A third aspect of universality of this data set has been identified by Argon \cite{Argon2013}. Introducing an appropriate normalization of the $(T/T_g)^{2/3}$ scaling relation, he has demonstrated that the athermal threshold stress $\hat{\tau} \cong 0.035 \, G$ also manifests universal character. Moreover, he pointed out that all metallic glasses that can be idealized as hard-sphere structures, regardless of their packing in various forms of short-to-medium-range order, have a rather universal plastic response in their yield behavior. 

\begin{figure}[htb!]%
\includegraphics[width=\columnwidth]{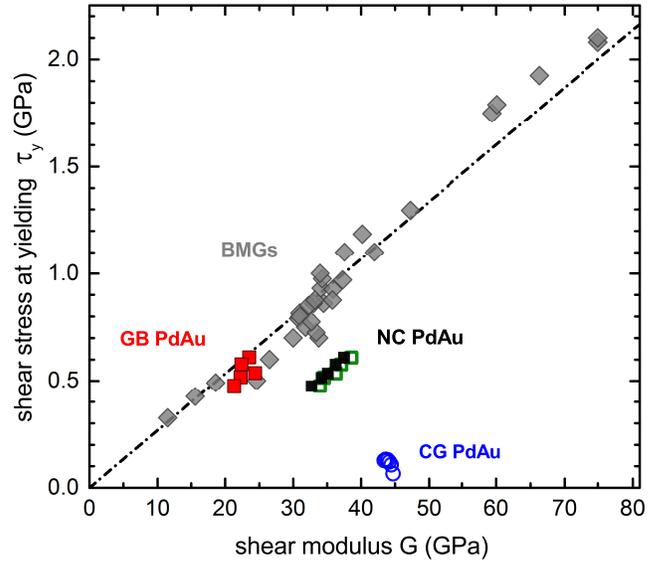}%
\caption{(Color online) Shear stress at yielding as a function of room temperature shear modulus. \textit{Open blue circles}: coarse grained Pd-Au, \textit{black squares}: NC Pd-Au samples related to their shear moduli not corrected for porosity, \textit{open green squares}: same samples related to shear moduli corrected for porosity, \textit{red squares}: green data points related to the effective shear moduli of grain boundaries. \textit{gray diamonds}: more than 30 different bulk metallic glasses from Johnson and Samwer\cite{Johnson2005}.}%
\label{fig5}%
\end{figure}

The available material parameters for the Pd-Au alloys enable to plot shear yield stress versus room-temperature shear modulus to reveal whether agreement or conflict prevails related to the universal behavior $\tau_y \approx 0.0267 \, G$ seen in BMGs. For comparison, we display data of shear stress at yielding versus room temperature shear modulus for a variety of metallic glasses taken from Johnson et al. \cite{Johnson2005} together with our data from CG and NC Pd-Au alloys in Fig. \ref{fig5}. We note that the data points of the CG specimens are certainly not related in any aspect to the metallic glasses since the latter cannot sustain the formation of dislocations or other soliton-like defects. By contrast, the data points of the nanoscale microstructures ($D \approx \unit[10]{nm}$) approach the slope of the universal behavior of BMGs (see Appendix for more details) but otherwise are shifted to shear modulus values being too large. A rationale that may explain this remaining discrepancy is the following. 

We have recently shown \cite{Grewer2011} that the shear modulus of GBs in NC metals is reduced by about 30\% compared to the respective bulk value. It seems therefore plausible to assume that the activation of STs takes essentially place in the core region of GBs. When the shear resistance of GBs as a consequence controls the onset of yielding, it thus naturally follows that the measured shear yield stress should correlate with the shear modulus $G_\mathrm{{gb}}$ of GBs. 

Based on the ray approach of ultrasound propagation and assuming that time-of-flights through crystallites and GBs are additive, we can write for the ultrasound velocity $v_{\mathrm{gb}}$ in GBs \cite{Grewer2011} 
\begin{equation}
v_{\mathrm{gb}} = \frac{\beta \, v_{\mathrm{nc}} \, v_{\mathrm{x}}}{\left( \beta - 1 \right) v_{\mathrm{nc}} + v_{\mathrm{x}}}
\label{eq:vGB}
\end{equation}%
where $v_{\mathrm{x}}$ is the sound velocity in the related CG material and $v_{\mathrm{nc}}$ is the overall ultrasound velocity in NC specimens. The parameter $\beta$ defines the length share of GBs which is proportional to $\delta / D$, where $\delta$ is the GB thickness; explicit expressions for $\beta$ and $\delta$ are given in Appendix A. By exciting transversal sound waves, measuring $v_{\mathrm{x}}$ and $v_{\mathrm{nc}}$ and solving for $v_{\mathrm{gb}}$ according to Eq.~\ref{eq:vGB}, it is straightforward to determine 
$G_\mathrm{{gb}} = \rho_\mathrm{{gb}} v_\mathrm{{gb}}^2$, where $\rho_\mathrm{{gb}}$ is the GB-density. 
Using a rule of mixture approach $\rho_{\mathrm{nc}} = (1-\alpha)\rho_{\mathrm{x}} + \alpha \rho_\mathrm{{gb}}$, we can solve for $\rho_\mathrm{{gb}}$, where $\rho_{\mathrm{nc}}$ represents the density of the NC Pd-Au alloys, $\rho_{\mathrm{x}}$ is set to the known bulk densities, and the parameter $\alpha$ represents the volume fraction of GBs; for more details we refer to Appendix A. 

Alluding to the rationale given above, it is recommended that we correlate shear yield stresses with $G_\mathrm{{gb}}$ instead of $G_\mathrm{{nc}}$. As shown in Fig. \ref{fig5}, the renormalization of $\tau_y$ with $G_\mathrm{{gb}}$ shifts the data points now right onto the straight line manifesting the universal behavior of BMGs. This evidence not only suggests that STs are dominant carriers of plastic strain in NC Pd-Au alloys but also implies that strain propagation dominantly takes place at/along GBs. Nevertheless, to make deformation happen in a compatible manner, other deformation mechanisms that support strain accommodation and dissipate local stress concentrations should coexist. Not least to avoid catastrophic failure appearing in the early stage of plastic material response. Likewise, using molecular dynamics simulations, Rupert \cite{Rupert2014} found that the yield strength for a broad variety of Cu-based NC alloys with $D = \unit[5]{nm}$ was linearly related to the Young's modulus of those samples in agreement with experimental work on metallic glasses by Takeuchi et al. \cite{Inoue2011}. Rupert pointed out that this behavior of NC metals manifests collective GB deformation physics reminiscent of amorphous metal deformation physics. 

The conclusions given above are in line with a detailed analysis of the microstructural evolution of NC $\mathrm{Pd}_{90} \mathrm{Au}_{10}$ during in situ deformation and high-energy X-ray microbeam diffraction \cite{Grewer2014arxiv}. The central evidence suggests that strain propagation in the so-called microplastic regime is solely due to linear elasticity and STs. The latter carry about $\unit[70]{\%}$ of the overall strain at the onset of yielding. %, which at the onset of yielding carry about 70 \% of the overall strain. 
Beyond yielding, dislocation activity and stress driven GB migration accompany STs but their respective share is on the order of 10\% only, just as the share of linear elasticity. Moreover, by investigating the stress-dependence of the free energy of activation, $\Delta G (\tau) $, related to inelastic deformation of NC $\mathrm{Pd}_{90} \mathrm{Au}_{10}$ alloys, Grewer et al. \cite{Grewer2014} found that the barrier height $\Delta G$ exhibits universal scaling behavior $\Delta G \propto \Delta \tau^{3/2}$, where $\Delta \tau$ is a residual load \cite{Maloney2006, Chattoraj2010}. They have also shown that this scaling behavior leads to a generalization of the universal $T^{2/3}$ temperature dependence of plastic yielding in metallic glasses. From the functional form of $\Delta G = \Delta G (\tau) $, the athermal threshold stress $\hat{\tau} = \tau (\Delta G = 0) =\unit [1.2]{GPa}$, representing the maximal shear resistance as $T \rightarrow \unit[0]{K}$, has also been determined \cite{Grewer2014}. The ratio $\hat{\tau}/G_{\mathrm{nc}} = 0.033$ compares favorably with Argon's universal relation $\hat{\tau} \cong 0.035 \, G$. When $T \rightarrow \unit[0]{K} $, we expect the thermally activated GB-mediated deformation modes to become frozen out. Therefore, we use the shear modulus $G_{\mathrm{nc}}$ to normalize $\hat{\tau}$ of NC $\mathrm{Pd}_{90} \mathrm{Au}_{10}$. Overall, it emerges that the yield (inelastic flow) behavior of NC $\mathrm{Pd}_{90} \mathrm{Au}_{10}$ alloys in the limit of small grain sizes agrees remarkably well with the three distinct aspects of the universal yield behavior of BMGs. It remains to be verified that this is true for the whole composition range. 

% Conclusion %%%%%%%%%%%%%%%%%%%%%
\section{Conclusions} %%%%%%%%%%%%%%%%%%%%
Studying solid solution effects on the strength of NC Pd-Au alloys, we present compelling evidence that the deformation physics of NC metals at the low end of the nanoscale ($D \lesssim \unit[10]{nm}$) is reminiscent of the deformation behavior of metallic glasses. In particular, it could be verified that the universal yield behavior of metallic glasses, i.e. the strictly linear relation between shear yield stress and shear modulus, is also obeyed for NC $ \mathrm{Pd}_{\mathrm{100-x}} \mathrm{Au}_{\mathrm{x}}$ alloys ($0 \leq \mathrm{x} < \unit[50]{at. \%}$). 

Moreover, we have shown that the predictions from dislocation-based models and theories related to solid solution effects on strength (hardening, softening) are violated. The general notion that twin- and stacking fault formation probabilities increase with decreasing ratio $\Lambda = \gamma_{\mathrm{isf}}/\gamma_{\mathrm{usf}}$ of intrinsic to unstable stacking fault energy \cite{Jin2011, VanSwygenhoven2004}, where $\Lambda$ is decreasing from $\approx 0.75$ for pure Pd to $\approx0.30$ in case of pure Au (see inset Fig. \ref{fig3}), is contradicted here. Altogether, this evidence also casts doubts on the applicability of traditional concepts of work or strain hardening in NC alloys at $D < \unit[10]{nm}$. As in metallic glasses, it would be desirable to study a variety of different material systems to further validate the findings discussed above. In the end, we would like to understand which atomic-scale feature(s) makes GBs propagating strain via shear transformations (STs), the generic flow defect in metallic glasses, but nevertheless avoiding serrated flow behavior and eventually runaway shear band formation. The identified alliance between metallic glasses and NC alloys may open an avenue to create ultrastrong, plastically deformable and catastrophic failure preventing alloys.

% ACKNOWLEDGEMENT %%%%%%%%%%%%%%%%%%%%%
\begin{acknowledgments} %%%%%%%%%%%%%%%%%%%%
The authors acknowledge the financial support of the Deutsche Forschungsgemeinschaft (FOR714 and 385/18-1) and fruitful discussions with T.J. Rupert on solid solution strengthening in nanocrystalline materials.
\end{acknowledgments}

%%%%%%%%%%%%%%%%%%%%%%%%%%%%%%%%%%%%%%%%%%%%%%%%%%%%%%%%%%%%%%%%%%%%%%%%%%%%%
% APPENDIX %%%%%%%%%%%%%%%%%%%%%%%%%%%%%%%%%%%%%%%%%%%%%%%%%%%%%%%%%%%%%%%%%%
\appendix %%%%%%%%%%%%%%%%%%%%%%%%%%%%%%%%%%%%%%%%%%%%%%%%%%%%%%%%%%%%%%%%%%%
\section{Determination of GB shear softening}

In what follows, we present the full set of equations, additional experimental- and literature data as well as assumptions and approximations that have been made to arrive at the conclusions drawn from Fig. 5.

\begin{figure}[htb!]%
\centering
\includegraphics[width=1\columnwidth]{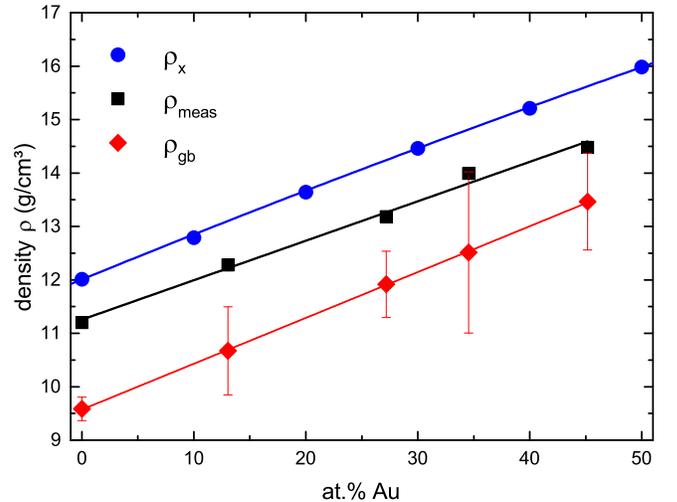}
\caption{(Color online) Experimentally determined density of NC Pd-Au, $\rho_{\mathrm{meas}}$ (error bars within symbol size), which includes porosity. For comparison, the theoretical crystal density $\rho_{\mathrm{x}}$ and grain boundary density $\rho_{\mathrm{gb}}$ derived from $\rho_{\mathrm{meas}}$ corrected for porosity are shown.}%
\label{fig:density}
\end{figure}

In Fig. \ref{fig:density}, we display the crystalline bulk density $\rho_{\mathrm{x}}$, which has been computed based on the continuous increase of lattice parameter with increasing Au-concentration and the associated change of the fcc unit cell volume of the continuous miscible Pd-Au alloy system. We also show the measured density $\rho_{\mathrm{meas}}$ of the as-prepared NC Pd-Au alloy specimen. 
The density deficit $\rho_{\mathrm{x}} -\rho_{\mathrm{meas}} $ of as-prepared specimens is related to porosity $P$, entailed by processing, as well as excess volume $\langle V_{\mathrm{ex}} \rangle $ stored in the core region of grain boundaries. 

The latter quantity is, in the spirit of Gibbsian excess quantities, defined as $\langle V_{\mathrm{ex}} \rangle := (V_{\mathrm{nc}} - V_{\mathrm{x}})/ A_{\mathrm{GB}} \equiv e $ and therefore has the dimension of a length; $V_{\mathrm{x}}$ represents the same amount of material as contained in $V_{\mathrm{nc}}$ but constituting a homogenous crystalline phase (reference state) characterized by $A / V_{\mathrm{x}} \rightarrow 0$, where $A$ represents surface and/or interface (grain boundary) area. Experimental and theoretical (computer simulations) values for $e$ are given in the pertinent literature \cite{Oberdorfer2014, Steyskal2012, Birringer1995}. %[cite Würschum acta 2014 and recent PRL and our Phil mag lett]. 
For NC Pd, we find values of $e$ for as prepared specimens in the range $\unit[0.06]{nm} < e < \unit[0.14]{nm}$; structurally relaxed specimens experience a roughly 50\% decrease of $e$. Since we compare as-prepared NC samples with as cast - neither annealed, aged nor rejuvenated - metallic glasses, it is in order to also utilize as-prepared values as input parameter for further analysis; in what follows, we employ the mean value $e = \unit[0.10]{nm}$. In case of Au, we find values of 
$e \approx \unit[0.01]{nm}$ for fully relaxed grain boundaries \cite{Merkle1992, Buckett1994}. In analogy to NC Pd, we assume that the excess volume of unrelaxed Au grain-boundaries amounts to $e \approx \unit[0.02]{nm}$, and we further assume, in lack of available data, that $e$ decreases linearly with increasing Au-concentration to eventually approach the pure Au-value.% We later show that these assumptions will not critically influence our central conclusion (\textbf{do we?})

Porosity is defined as $P := 1 - (\rho_{\mathrm{meas}}/\rho_{\mathrm{nc}}) $ where $\rho_{\mathrm{nc}}$ is the density of pore-free but excess volume carrying NC material. Alternatively, the measured density is given as $\rho_{\mathrm{meas}} = P\rho_{\mathrm{void}} + (1-P)\rho_{\mathrm{nc}} $ which simplifies to $\rho_{\mathrm{meas}} = (1-P)\rho_{\mathrm{nc}}$ as $\rho_{\mathrm{void}} \approx 0$. Since NC metals in the limit of small grain sizes can be treated as statistically homogenous and isotropic objects, we use a rule of mixture approach to express $\rho_{\mathrm{nc}}$ in terms of $\rho_{\mathrm{x}}$ and $\rho_{\mathrm{gb}}$ where the latter quantity denotes the grain-boundary density. With the volume fraction $\alpha$ of 
grain boundaries (interface), we then obtain
\begin{equation}
\label{eq:densityNC}
\rho_{\mathrm{nc}} = \alpha\rho_{\mathrm{gb}} + (1-\alpha)\rho_{\mathrm{x}}.
\end{equation}
The ansatz we use to estimate $\alpha = A_{\mathrm{gb}} \delta / V = 2 \delta / \langle L \rangle_{\mathrm{area}} $ relies upon the stereological identity $A/V = 2 / \langle L \rangle_{\mathrm{area}}$, where $A/V$ represents the interface area per unit volume of crystal and $\langle L \rangle_{\mathrm{area}}$ denotes the 
area-weighted average column length of crystal. In reference \cite{Krill1998}, we show that $\langle L \rangle_{\mathrm{area}}$ is related to the experimentally (XRD) extracted grain size via $\langle D \rangle_{\mathrm{vol}} = \frac{3}{2} \langle L \rangle_{\mathrm{area}} \exp{\{(\ln \sigma)^2\}}$, where $\sigma$ measures the width of the grain size distribution function, which is also deducible from peak profile analysis of x-ray diffraction patterns \cite{Krill1998}. The symbol $\delta$ describes the average structural width of grain boundaries and enters $\rho_{\mathrm{gb}} := m / A_{\mathrm{gb}} \cdotp \delta$. Assuming that low hkl-indexed lattice planes abut the grain boundary, it is practical and in the spirit of the structural unit model of GBs \cite{Sutton1983, Howe1997} to write $\delta \approx 3 (a/\sqrt{3}) +e$ where $a$ denotes the lattice parameter. Since $e$ as well as $a$ depend on the Au-concentration, the structural width is also concentration dependent. However, we compute that the increase of $a$ with increasing Au-concentration is basically compensated by the decrease of $e$, so it is straightforward to accept 
$\delta \approx \unit[(0.76 \pm 0.01)]{nm}$ being constant, where $a_{\mathrm{Pd}} = \unit[0.389]{nm}$ has been used. We eliminate the mass $m$ by looking into the ratio $\rho_{\mathrm{gb}} / \rho_{\mathrm{x}} = 1 /(1+e/(\sqrt{3}a))$. 

As a result, $\rho_{\mathrm{meas}}$ becomes a function of $P, D_{\mathrm{vol}}$ and $ \sigma $ when $\delta$ is treated constant. We can solve now for $P$ to retrieve the composition-dependent porosity related to the specimens shown in Fig. \ref{fig23}. With these results (see table \ref{table1}), we are able to also display the shear modulus 
$G_{\mathrm{nc-corr}} = \rho_{\mathrm{nc}}\, v_{\mathrm{s, nc}}^2 = \rho_{\mathrm{meas}} / (1-P)\, v_{\mathrm{s, nc}}^2 $, which is corrected for porosity. Overall, we find that such small amounts of porosity have rather little influence on the shear modulus when compared with the difference between coarse-grained bulk and nanocrystalline material (see Fig. \ref{fig23} (b)). 

\begingroup
%\squeezetable
\begin{table*}[t]
 \caption{Parameters: 
gold concentration in PdAu, lattice parameter $a$, width of the
log-normal crystallite size distribution $\sigma$, GB volume fraction $\alpha$, GB length fraction $\beta$, Porosity $P$, transversal sound velocity $v_{\mathrm{s,nc}}$, measured density $\rho_{\mathrm{meas}}$, grain size $D_{\mathrm{vol}}$, GB-density $\rho_{\mathrm{gb}}$, GB-shear modulus $G_{\mathrm{gb}}$
}\label{table1}
\begin{ruledtabular}
\begin{tabular}{lrrrrr}
at.\% Au & 0.00	& 13.05 &	27.18 &	34.55 &	45.17 \\
a (pm)  & $ 388.5 \pm 0.2  $ & $	391.\pm 0.1  $ & $	393.5 \pm 0.1  $ & $	395.3 \pm 0.2  $ & $	397.1 \pm 0.2 $ \\
$\sigma$  & $ 1.70 $ & $	1.70	$ & $ 1.71	$ & $ 1.65 $ & $	1.58 $\\
$\alpha$  & $ 0.23 \pm 0.01  $ & $	0.23 \pm 0.03  $ & $	0.28 \pm 0.03  $ & $	0.29 \pm 0.04  $ & $	0.28 \pm 0.03 $ \\
$\beta$  & $ 0.29	\pm 0.03  $ & $ 0.30 \pm 0.05 	$ & $ 0.38 \pm 0.06 	$ & $ 0.37 \pm 0.07  $ & $	0.34 \pm 0.05 $ \\
$P$ (\%)  & $ 2.58 $ & $	2.13 $ & $ 3.18 $ & $ 1.04 $ & $	3.56 $ \\
$v_{\mathrm{s,nc}}$ (m/s)  & $ 1566 \pm 50  $ & $	1447 \pm 71  $ & $	1430 \pm 58  $ & $	1332 \pm 78  $ & $	1257 \pm 72 $ \\
$\rho_{\mathrm{meas}}$ (g/cm$^3$)  & $ 11.20 $ & $	12.278 \pm 0.006 $ & $	13.184 \pm 0.007$ & $	13.992 \pm 0.003 $ & $	14.479 \pm 0.003 $\\
$D_{\mathrm{vol}}$ (nm)  & $ 10.5 \pm 1.1 	$ & $ 10.0 \pm 2.0 	$ & $ 7.7 \pm 1.4  $ & $	7.2 \pm 1.8  $ & $ 7.0 \pm 1.2 $ \\
$\rho_{\mathrm{gb}}$ (g/cm$^3$)  & $ 9.6 \pm 0.2  $ & $ 10.7 \pm 0.8  $ & $	11.9 \pm 0.6  $ & $	12.5 \pm 1.5  $ & $	13.4 \pm 0.9 $ \\
$G_{\mathrm{gb}}$ (GPa)  & $ 24 \pm 2  $ & $ 22 \pm 3  $ & $	24 \pm 2  $ & $	22 \pm 4  $ & $	21 \pm 3  $
%a [pm] & 388.5 (2) &	391.(1) &	393.5 (1) &	395.3 (2) &	397.1 (2) \\
%$\sigma$ & 1.70 &	1.70	& 1.71	& 1.65 &	1.58 \\
%$\alpha$ & 0.23 (1) &	0.23 (3) &	0.28 (3) &	0.29 (4) &	0.28 (3) \\
%$\beta$ & 0.29	(3) & 0.30 (5)	& 0.38 (6)	& 0.37 (7) &	0.34 (5) \\
%$P$ [\%] & 2.58 &	2.13 & 3.18 & 1.04 &	3.56 \\
%$v_{\mathrm{s,nc}}$ [m/s] & 1566 (50) &	1447 (71) &	1430 (58) &	1332 (78) &	1257 (72) \\
%$\rho_{\mathrm{meas}}$ [g/cm$^3$] & 11.20 &	12.28 &	13.18 &	13.99 &	14.48 \\
%$D_{\mathrm{vol}}$ [nm] & 10.5 (1.1)	& 10.0 (2.0)	& 7.7 (1.4) &	7.2 (1.8) & 7.0 (1.2) \\
%$\rho_{\mathrm{gb}}$ [g/cm$^3$] & 9.6 (2) & 10.7 (8) &	11.9 (6) &	12.5 (1.5) &	13.4 (9) \\
%$G_{\mathrm{gb}}$ [GPa] & 24 (2) & 22 (3) &	24 (2) &	22 (4) &	21 (3) 
\end{tabular}
\end{ruledtabular}
\end{table*}
\endgroup

The quantity of interest is the shear modulus of grain boundaries $G_{\mathrm{gb}}$ that is formally defined as:
\begin{equation}
\label{eq:gbshearmod}
G_{\mathrm{gb}} = \rho_{\mathrm{gb}}\, v_{\mathrm{s, gb}}^2, 
\end{equation}
where $\rho_{\mathrm{gb}}$ is a known quantity as discussed above. An expression for $v_{\mathrm{gb}}$, generally applicable to transverse or longitudinal sound velocity, has been derived by Grewer et al.\cite{Grewer2011} based on the assumption that running times of sound waves across crystalline and grain boundary phase add up. Eq. \ref{eq:vGB} of the manuscript represents this expression which we repeat here for the sake of clarity: 
\begin{equation}
\label{eq:gbvelocity}
v_{\mathrm{gb}} = \frac{\beta \, v_{\mathrm{nc}} \, v_{\mathrm{x}}}{\left( \beta - 1 \right) v_{\mathrm{nc}} + v_{\mathrm{x}}}.
\end{equation}

\begin{figure}%[t!]%
\includegraphics[width=\columnwidth]{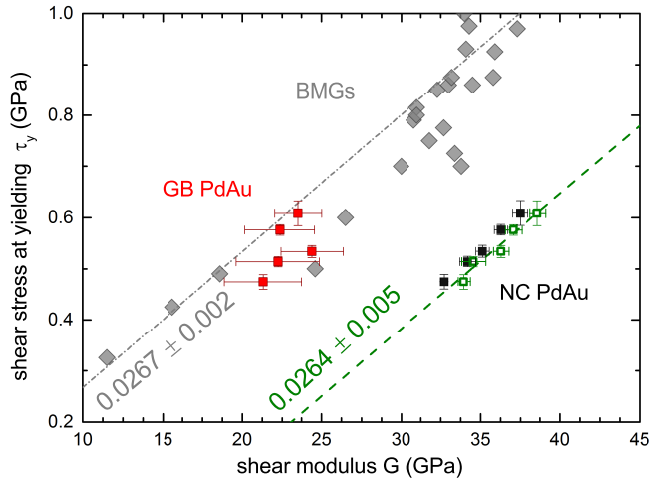}%
\caption{(Color online) Shear stress at yielding as a function of room temperature shear modulus. \textit{Black squares}: NC Pd-Au samples related to their shear moduli not corrected for porosity, \textit{open green squares}: same samples related to shear moduli corrected for porosity, \textit{red squares}: green data points related to the effective shear moduli of grain boundaries. \textit{Gray diamonds}: more than 30 different bulk metallic glasses from Johnson and Samwer\cite{Johnson2005}. \textit{Dashed dotted gray line}: universal yield behavior of BMGs, \textit{dashed green line}: linear fit to green data points.}%
\label{fig:App2}%
\end{figure}

Apart from the experimentally determined values for $v_{\mathrm{s,x}}$ and $v_{\mathrm{s, nc}}$, the length share of grain boundaries \cite{Grewer2011}, $\beta = 4\delta /(4\delta+3\langle L \rangle_{\mathrm{area}} \exp{\{ -2 \ln (\sigma)^2 \}}) %= 2\delta / (\delta + \exp\{\ln \langle D \rangle_{\mathrm{vol}} -3 (\ln \sigma )^2\} )
$, enters $v_{\mathrm{s, gb}}$ but is also a function of known quantities. 
Putting all this information (see table \ref{table1}) into Eq. \ref{eq:gbshearmod} finally yields the reevaluated data points displayed in Fig. 5. We are aware of a computer simulation study \cite{Bachruin2014} addressing the effect of porosity on the elastic and yield behavior of NC Pd. Our observed slight decrease of moduli is consistent with their results. Different from our observations, they identify the onset of plasticity at much larger stresses, on the order of $\unit[5]{GPa}$, which has been associated with dislocation nucleation at stress concentrators. It is also found that porosity entails a linear but slight decrease of the onset of yielding. We refrain from correcting our shear yield stress values for porosity by adapting their data because it is a priori not obvious how different deformation mechanisms are affected by porosity and how this is influencing the yield behavior. In any case, we would however expect that our yield stress values would shift to larger values if it were possible to come up with a valid correction procedure for porosity. As a final result, Fig. \ref{fig:App2} shows with magnified scale to which precision the bulk metallic glasses, our corrected and uncorrected data points follow the universal behavior represented by the dashed-dotted straight line. The observation of basically identical slopes of BMGs and NC Pd-Au lends additional support to our contention that the deformation physics of NC metals at the low end of the nanoscale is reminiscent of the deformation behavior of metallic glasses.

\end{document}